\documentclass[conference,twocolumn]{IEEEtran}

\usepackage{amsmath}
\usepackage{amsthm}
\usepackage{amssymb}
\usepackage{mathtools}
\usepackage{rotating}
\usepackage{enumitem}
\usepackage{tcolorbox}
\usepackage{xcolor}
\usepackage{tikz}

\newsavebox{\tempbox}

\DeclareMathAlphabet\mbcf{OMS}{cmsy}{b}{n}

\newcommand{\logt}{\log_{2}}
\newcommand{\dx}{\mathrm{d}}

\newcommand{\defn}{\triangleq}

\newcommand{\mcf}[1]{\mathcal{#1}}

\newcommand{\set}[1]{\left\{ #1 \right\}}

\newenvironment{proofsketch}{%
  \proof}{\endproof}

\newtheorem{theorem}{Theorem}
\newtheorem{define}[theorem]{Definition}

\newtheorem{lemma}[theorem]{Lemma}
\newtheorem{cor}[theorem]{Corollary}

\newtheorem{remark}[theorem]{Remark}

\let\oldremark\remark
\renewcommand{\remark}{\oldremark\normalfont}

\IEEEoverridecommandlockouts

\usepackage{etoolbox}

\providetoggle{arxiv}
\settoggle{arxiv}{true}

\begin{document}

\title{{An information theoretic model for summarization, and some basic results}
\thanks{Research was sponsored by the Army Research Laboratory and was accomplished under Cooperative Agreement Number W911NF-09-2-0053 (the ARL Network Science CTA). The views and conclusions contained in this document are those of the authors and should not be interpreted as representing the official policies, either expressed or implied, of the Army Research Laboratory or the U.S. Government. The U.S. Government is authorized to reproduce and distribute reprints for Government purposes notwithstanding any copyright notation here on.}}

\author{\IEEEauthorblockN{Eric Graves}
\IEEEauthorblockA{Army Research Lab\\
CISD \\
Adelphi, MD 20783\\
ericsgraves@gmail.com}
\and
\IEEEauthorblockN{Qiang Ning}
\IEEEauthorblockA{University of Illinois at Urbana-Champaign\\
Department of Computer Science\\
Urbana, IL 61801\\
qning2@illinois.edu }
\and
\IEEEauthorblockN{Prithwish Basu}
\IEEEauthorblockA{Raytheon BBN Technologies\\
Networking and Cyber Technologies\\
Cambridge, MA 02138\\
prithwish.basu@raytheon.com }}

\maketitle

\begin{abstract}
A basic information theoretic model for summarization is formulated. 
Here summarization is considered as the process of taking a report of $v$ binary objects, and producing from it a $j$ element subset that captures most of the important features of the original report, with importance being defined via an arbitrary set function endemic to the model.
The loss of information is then measured by a weight average of variational distances, which we term the semantic loss.

Our results include both cases where the probability distribution generating the $v$-length reports are known and unknown.
In   the   case   where   it   is   known, our  results  demonstrate  how  to  construct  summarizers  which minimize  the  semantic  loss. 
For  the  case  where  the  probability distribution is unknown, we show how to construct summarizers whose  semantic  loss  when  averaged  uniformly  over  all  possible distribution converges to the minimum.

\end{abstract}

\section{Introduction}

For a concrete example of how we shall define information summarization, consider the following weather report.
\begin{center}
\begin{tabular}{ l | c }
Phenomena &  \\
\hline
High winds & \checkmark \\
High UV index & \checkmark \\ 
Heavy Rain & \checkmark \\
Snow & \\
Low visibility & \checkmark \\
Smog & \\
Typhoon & \checkmark 
\end{tabular}
\end{center}
Our stance is that such a report is overly detailed, and wish to design a system that produces summaries such as the following.
\begin{center}
\begin{tabular}{ l | c c l | r }
Phenomena &  & or & Phenomena & \\
\cline{1-2} \cline{4-5}
High UV index & \checkmark & \hspace{10pt} & High UV index & \checkmark \\ 
Typhoon & \checkmark & \hspace{10pt} & Smog &  \\
 \multicolumn{1}{r}{} & & & Typhoon & \checkmark
\end{tabular}
\end{center}
In this example it is important to note a typhoon already implies high winds, heavy rain and low visibility, and heavily implies the absence of snow. 
At the same time, the presence of a typhoon does not generally indicate a high UV index or the lack of smog; these events should still be reported. 

In the abstract, the goal of summarization is to reduce the dimension of data, without excessive loss of ``information.''
This abstract notion is very similar to that of compression and rate-distortion theory~\cite[Chapter~5,~10]{CT}; summarization is distinguished in two important ways.
First, unlike compression and rate distortion which feature both an encoder to construct the efficient representation and a decoder to interpret this representation, a summarizer only has a single element, the output of which should be ready for immediate consumption.
%Second, but whose need is predicated by the first, a performance measure of the summarizer's output needs to be independent of the summarizer algorithm\footnote{Consider compression, where the probability of error }. 
%Indeed, consider again the weather report, where a summarizer can produce $\left(\begin{matrix} 7 \\ 3 \end{matrix} \right) \cdot 2^3 = 280$ possible $3$-phenomena summaries\footnote{In theory a summary does not necessarily need to be truthful to the report, although in practice one hopes it would be.} to summarize $2^7$ possible reports. 
%Thus, a summarizer designed under the assumption that the end user knows the summarization algorithm would allow for solutions that convey all information, but are completely unintelligible by the average person.
Second, we must take into account the importance of the underlying information, as opposed to simply the likelihood. 
For instance, smog may be less likely than a typhoon, but the typhoon is more essential to include given that both occur.

Despite similarities to established concepts in information theory, to the best of our knowledge, summarization has never been considered from the information theoretic perspective.
%\footnote{An all together different notion of ``summarization'' was studied previously from an information theoretic perspective by Barron et al.~\cite{barron1998minimum}. There, summarization was about select best parameters to estimate an underlying model.}.
Instead most of the literature exists within the natural language processing community and the machine learning community, see \cite{lin06information, lin2010multi, lin2012learning} and references there within. 
The approach of these communities is to directly engage the general problem, searching for efficient practical solutions which provide empirically good results\footnote{For discussion on this viewpoint, see Simeone \cite[Chapter~1]{simeone2017brief}.}.
This differs from a traditional information theoretic approach, where a simplified model is established and analyzed in order to determine fundamental limits of the operational parameters, and to gain insight into how to achieve those limits\footnote{For discussion on this viewpoint, see Han~\cite[Preface]{han2003}.}.  

To simplify this model, we shall make the following assumptions.
First, the data to be summarized is a length-$v$ binary sequence, which has an arbitrary (not necessarily independent) probability distribution relating each symbol.
While the probability distribution over a length-$v$ binary sequence is arbitrary, every length-$v$ sequence the summarizer observes is independent and identically distributed.
Second, we assume that the summarizers output needs be ``extractive,'' meaning that the summarizer can only produce a subset of what is input, as in the weather example.
Finally, we assume the existence of an arbitrary set function that can be used to measure the ``semantic information'' of a random variable.
This last assumption will be further justified in Section~\ref{sec:pssm:model} via example, but it is worth mentioning that, as shown by Yeung~\cite{yeung1991new}, Shannon's measure of (nonsemantic) information (entropy) has such a representation.
Spurred by this, Lin and Bilmes~\cite{lin2010multi, lin2012learning} have recently argued for the use of submodular functions in an effort to axiomatically define a notion of semantic information. 
Regardless, we will make no other assumptions on this function other than existence and that it is finite and positive.

%\cite{lin06information}
%\cite{lin2010multi}
%\cite{lin2012learning}

%\cite{carbonell1998use} 

\section{Notation}

\emph{Random variables} (RV(s)) will be written in upper case, constants in lower case, and sets in calligraphic font. For example $X$ can take on value $x$ from $\mcf{X}.$
A $n$-length sequence of random variables, variables or sets will be denoted with the power $n$, such as $X^n$. 
Among sets $\mcf{P}(\mcf{X}|\mcf{Y})$ will hold special meaning as the set of conditional probability distributions on the set $\mcf{X},$ when given a value from the set $\mcf{Y}$. That is, if $p \in \mcf{P}(\mcf{X}|\mcf{Y})$, then $p_{y}(x) \in [0,1]$ and $\sum_{x \in \mcf{X}} p_{y}(x) = 1$ for all $x \in \mcf{X},~y \in \mcf{Y}$.
For convenience, given $\mcf{ W} \subset \mcf{X}$ then $\mcf{P}(\mcf{W})\subset \mcf{P}(\mcf{X})$, where only symbols in $\mcf{W}$ have non-zero probability.
The symbol $\sim$ will be used to relate probability distributions and random variables. For example if $X \sim p(x)$ and $Y \sim q_{X}(y)$, for some $q \in \mcf{P}(\mcf{Y}|\mcf{X})$ and $p \in \mcf{P}(\mcf{X})$, then $\Pr \left( X =x , Y=y \right) = q_{x}(y) p(x)$.
When a set is used in a probability distribution, such as $p(\mcf{\hat X})$ for some $\mcf{\hat X} \subset \mcf{X}$ and $p \in \mcf{P}(\mcf{X}),$ it means $\sum_{x\in \mcf{\hat X}} p(x).$

%For any convex function $f: \mathbb{R} \rightarrow \mathbb{R}$, where $f(1) = 0$,
%$$\mathbb{D}_{f}(p||q) \defn \sum_{x \in 2^{\mcf{V}}} q(x) f\left( \frac{p(x)}{q(x)} \right)  $$
%denotes the $f$-divergence (for more see~\cite[Pg.~447]{csiszar2004information}). In particular  $$D_{t \ln t}(p||q) = \sum_{x \in \mcf{X}} p(x) \ln \frac{p(x)}{q(x)},$$
%is the information (or KL) divergence between $p$ and $q$. 

We shall use $\pi^{(x^n)}$ to denote the empirical distribution of a series $x^n$, for example $\pi^{(0,1,1,1)}(0) = \frac{1}{4}$ while $\pi^{(0,1,1,1)}(1) = \frac{3}{4}.$
The set $\mcf{T}^n_{(x^n)}$ denotes the set of $n$-length sequences with the same empirical distribution as $x^n$, that is $\mcf{T}^n_{(x^n)} \defn \{ \hat x^n : \pi^{(\hat x^n)} = \pi^{(x^n)}\}.$ 
It will be important to note that 
\begin{align*}
|\mcf{T}^n_{(x^n)}| %&\defn \left| \set{ \tilde x^n \in \mcf{X}^n : \pi^{(\tilde x^n)} = \pi^{(x^n)} }\right| \\
&= \left( \begin{matrix}n \\ n\pi^{(x^n)}(1),\dots, n \pi^{(x^n)}(|\mcf{X}|) \end{matrix}\right).
\end{align*}
Next $\mcf{P}_{n}(\mcf{X})$ denotes the set of valid empirical distributions for  $n$-length sequences of symbols from $\mcf{X}$. Again it is important to note
\begin{align*}
|\mcf{P}_{n}(\mcf{X})| %&= \left|\set{q \in \mcf{P}(\mcf{X}) : \exists  x^n \in \mcf{X}^n \text{ for which } q = \pi^{(x^n)} } \right|\\
&= \left( \begin{matrix}n + |\mcf{X}| - 1 \\ |\mcf{X}|- 1\end{matrix}\right) .
\end{align*}
%When a set is raised to a parenthetical set, it is to represent the set formed from the base set by combining all elements of the superscript set into a single element. 
%For instance $\{1,2,3,4,5\}^{(\{1,4,5\})} = \{ 2,3, \{1,4,5\} \}.$
%Likewise if a function (or probability distribution) is raised to a parenthetical set it denotes a new function of whose domain has been reformed using the raised set, and whose outputs are that of the original function except for the new super symbol whose output is the super position of the symbols. 
%That is, for a function $f$ with domain $\mcf{X}$ the function $f^{(\mcf{\hat X})}$ has domain $\mcf{X}^{(\mcf{\hat X})}$ and $$f^{(\mcf{\hat X})} (x') = \begin{cases}  f(x') &\text{ if } x' \in \mcf{X} - \mcf{\hat X} \\ \sum_{\tilde x \in \mcf{\hat X}} f(\tilde x) &\text{ if } x' = \mcf{\hat X} \end{cases} ,$$  for any $\mcf{\hat X} \subseteq \mcf{X}$.
$1_{\mcf{\hat X}}$ is the indicator function, $$ 1_{\mcf{\hat X}}(x) = \begin{cases} 1 & \text{ if } x \in \mcf{\hat X} \\ 0 & \text{ o.w. } \end{cases},$$
and $\mathbb{E}$ is the expected value operator, where the expectation is taken over all random variables.

\section{System Model and Justification}

%In Section~\ref{sec:pssm:model} we define the summarization model which we will analyze throughout the paper. 
%This model is quite general, and because of that, it may not be clear that it is an acceptable operational model for the problem.
%With that in mind, we shall show in Section~\ref{sec:mod ex} that the model can be specialized to two different models which are operationally relevant.

%\subsection{Model}
\label{sec:pssm:model}

\tikzstyle{circ} = [draw, fill=white, circle, node distance=1cm]

\tikzstyle{block} = [draw, fill=white, rectangle, 
    minimum height=30pt, minimum width=10pt, text centered]

\tikzstyle{bigblock} = [draw, thick, fill=orange, rectangle, 
    minimum height=30pt, minimum width=30pt, text centered]

\begin{figure}
    \centering
\begin{tikzpicture}
\node[block] (report) at (3,0) {$\begin{array}{c} \text{Current Report} \\ X \sim p(x), \\ p \in \mcf{P}(\mcf{X}) \end{array}$};
\node[bigblock] (summ) at (5.8,0){$\begin{array}{c} \text{Summarizer} \\ Y \sim  s_{X^n}(y), \\ s \in \mcf{P}( \mcf{Y} | \mcf{X}^{n}) \end{array}$};
\node[block] (thist) at (5.8,-1.75) {$\begin{array}{c} \text{Report History} \\  (X(2),\dots,X(n)) \sim \prod_{i=2}^{n} p(x(i)) \end{array}$};
\node[block] (user) at (8.85,0) {$\begin{array}{l} \text{User with} \\
\text{Semantic Weights:} \\ u: \mcf{X} \times 2^{\mcf{X}} \rightarrow \mathbb{R}^+ \end{array}$};
\draw[->,thick] (report.east) -- (summ.west) ;
\draw[->,thick] (thist.north) -- (summ.south);
\draw[->,thick] (summ.east) -- (user.west);
%\draw[->,thick] (dec.east) -- node[above]{$\hat M$} (9,0) ;
%\node (name) at (2,-2) {$\begin{array}{l} \text{    System}\\ \text{Encoder:} \mcf{M} \rightarrow \mbcf{X} \\ \text{Decoder:}   \mbcf{Y} \rightarrow \mcf{M}  \end{array}$ };
%\draw[->,thick] (name.north) .. controls (1.8,-1) .. (1,-1) .. controls (.5,-1) .. (enc.south);
%\node (cprop) at (5,-3) {$\begin{array}{r}  p_{\mbf{Y}|\mbf{X}}(\mbf{y}|\mbf{x}) = \prod_{i=1}^n p_{Y|X}(y_i|x_i) \end{array}$};
\end{tikzpicture}
 \caption{Model, with design elements highlighted orange.}
\label{fig:model}
\end{figure}
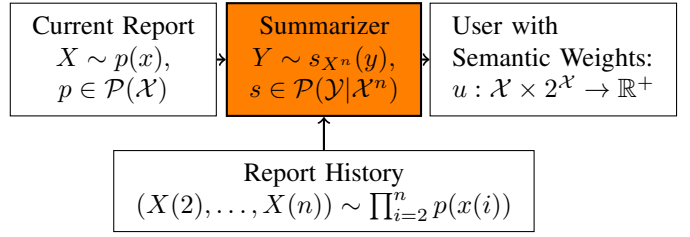

The objects to be summarized will be referred to as \emph{reports}.
A sequence of $n$-reports will be denoted by a sequence of RVs $X^n = (X(1),\dots, X(n)) \sim \prod_{i=1}^n p(x(i))$, where $p \in \mcf{P}(2^{\mcf{V}})$ and $\mcf{V}$ is the finite set of \emph{possible events}.
Without loss of generality we will assume $\mcf{V} = \set{1,\dots,v}$ for some positive integer $v$, and accordingly will refer to $X$ as a $v$-symbol binary sequence with $x_{j} = 1$ denoting possible event $j \in \mcf{V}$ occurring for report $x$. 
From here forward $\mcf{X} \defn 2^{\mcf{V}},$ or the power set of $\mcf{V}$, for convenience.

Although given $n$-reports the summarizer only needs to summarize $X(1)$, as shown in Figure~\ref{fig:sum_exp}; to summarize is to produce a subset of $j \in \mathbb{Z}^+$ possible events and indicate whether or not they each occurred.
Formally, the summarizer produces $Y = (\hat Y,\tilde Y)$, $\mcf{\hat Y} = \left( \begin{matrix} \mcf{V} \\ j \end{matrix}\right) $, $\mcf{\tilde Y}= \{0,1\}^j$, where $\hat Y$ is a subset containing $j$ possible events and $\tilde Y$ is the indication of whether or not the possible events in $\hat Y$ occurred.
Because the summarizer only needs to summarize the first report, we will refer to $X(1)$ as the \emph{current report}, and $X_2^{n}\defn (X(2),\dots, X(n))$ as the \emph{report history}.
Note that finding the optimal summary for $X(1)$ also finds the optimal summary algorithm for $X(k)$ for $k = \{2,\dots, n\}$ since they are identically distributed.

Notice that a summary does not necessarily provide all of the information about the report, and moreover there are multiple reports for which a given summary may be the representative. 
A specific summary $y$ may be the summary for any report in\footnote{We write $y \subset x$ if a summary $y$ could lead to $x.$
That is, if $\hat y = \{1,2,5\}$ then $y = (\hat y,\tilde y) \subset x$ if and only if $\tilde y = (x_1,x_2,x_5)$.}
$$\mcf{X}(y) \defn \{ x \in \mcf{X}: y \subset x \}.$$
Clearly for a given summarization algorithm, each $x \in \mcf{X}(y)$ does not necessarily generate summary $y$.

To relate the output of the summarizer to the input we introduce a conditional probability distribution called the \emph{summary interpretation},
$$i_{y}(x) \defn 1_{\mcf{X}(y)}(x) \frac{p(x) }{  p(\mcf{X}(y)) }.$$
The summary interpretation is equal to the probability of a report for a given summary when averaged uniformly over all possible summarizers.
In this way it represents an end user which has absolutely no information about the summarizing algorithm, but knows perfectly the distribution of what is being summarized.

Having related the report and the summary, our goal will be to produce summaries which capture most of the ``semantic information'' of the current report. 
In order to capture ``semantic information,'' included in the model will be a set function $u : \mcf{X} \times 2^{\mcf{X}} \rightarrow \mathbb{R}^+$ denoting \emph{semantic weights}.
The semantic weights assign to each pair of $x \in \mcf{X}$ and subset $\mcf{W}\subseteq \mcf{X}$ a number representative of the importance that the summary convey $x\in \mcf{W}$ (if $x\notin\mcf{W}$ then $u(x,\mcf{W}) = 0$). 
The motivation behind the semantic weights is best viewed by the weather example.
With each report there are a number of different possible implications, for instance, the report might imply extra layers of clothing are needed, or that it is a nice day to go outside, or that serious meteorological event is occurring. 
Each of these implications can be considered semantic information, as it has some intuitive meaning. 
Furthermore, each of these implications is valid for a large number of possible reports.
In that sense, each set $\mcf{W}$ is representative of some semantic meaning shared between the collective of reports in the set, and $u(x,\mcf{W})$ is represents how important this semantic meaning is to the report $x$.

\begin{figure}
\centering
\begin{tikzpicture}
  \node at(.5,1.5) {$v = $};
  \node at(.27,.5) {$x(1) = $};
  \node at(.5,-1.5) {$\tilde y =$};
  \node at(.5,-.5) {$\hat y = $};
\foreach \x/\xtext in {1/1,2/0,5/0 }  
  {
  \filldraw[fill=blue!20!white,thick, draw=black!50!black] (\x,0) rectangle (\x+1,1);
  \node at (\x+.5,1.5) {$\x$};
  \node at (\x+.5,.5) {$\xtext$};
  \node at (\x+.5,-1.5) {$\xtext$};
  \node at (\x+.5,-.5) {$\x$};
  }
 \foreach \x/\xtext in {3/0,4/1,6/0,7/1 }  
  {
  \filldraw[fill=white, draw=black!50!black] (\x,0) rectangle (\x+1,1);
  \node at (\x+.5,1.5) {$\x$};
  \node at (\x+.5,.5) {$\xtext$};
  }
\end{tikzpicture}
\caption{Summarizer input and output}
\label{fig:sum_exp}
\end{figure}
 
Having defined all aspects endemic to the model, we now move to discussing the operational parameters.
To aggregate and measure the performance of the summarizer, we shall use the semantic loss, which is the summarizer analog to a distortion criteria in rate distortion theory.
\begin{define} \label{def:sem_loss}
The \emph{semantic loss of $X$ to $Y$ with semantic weights $u: \mcf{X} \times 2^{\mcf{X}} \rightarrow \mathbb{R}^+$} is 
\begin{align}
\ell(X;Y | u) \hspace{-2pt}
&\defn  \hspace{-2pt}    \mathbb{E }\hspace{-3pt}\left[ \hspace{-1pt}\sum_{\mcf{W} \subseteq \mcf{X}} \inf_{q \in \mcf{P}(\mcf{W}|\mcf{X})} \hspace{-8pt} u(X, \mcf{W}) \hspace{-2pt} \sum_{x \in \mcf{X}}\frac{ \left| q_{X}(x) - i_{Y}(x) \right|}{2} \right]\hspace{-3pt}
\notag. %\\
%&= \mathbb{E} \left[  \sum_{ \mcf{W} \subseteq \mcf{X} }  u(X, \mcf{W})  \left( 1 - i_{Y}(\mcf{W}) \right) \right] \label{eq:def:convex_closure}.
\end{align}
\end{define}
%\begin{remark}
%We have chosen to present Equation~\eqref{eq:def:convex_closure} directly in the definition, because it will be helpful for discussion purposes. It is somewhat obvious; a proof of this statement can be found in Appendix~\ref{app:convex_closure}.
%\end{remark}
\noindent Consider the semantic loss when there is a single $\mcf{W}$ such that $u(x,\mcf{W}) \neq 0. $ 
In this case the semantic loss is the variational distance between the summary interpretation and the closest distribution such that only reports in $\mcf{W}$ occur.
Clearly, if only reports in $\mcf{W}$ were possible, given a particular summary, then this summary would losslessly convey that a report in $\mcf{W}$ occurred. 
Using an $f$-divergence (see~\cite[Chapter~4]{csiszar2004information}), namely variational distance, give us a well studied way to then measure the distance between the summary interpretation and the convex set of distributions which perfectly convey $\mcf{W}$. 
This distance is then averaged over all semantic meanings according to the semantic weights.

We conclude the section with a more formal definition of a summarizer. 
For the purpose of easily specifying operational parameters, we shall refer to a summarizer by the probability distribution relating the summary $Y$ and the reports $X^n$.
\begin{define}
For each $j \in \mathbb{Z}^+$ and $\delta \in \mathbb{R}^+$,
a \emph{summarizer} $s \in \mcf{P}(\mcf{Y}|\mcf{X}^n)$ has \emph{length} $j$ if $$\mcf{Y} = \left( \begin{matrix} \mcf{V} \\ j \end{matrix}\right) \times \{0,1\}^j$$ 
and has \emph{$\delta$ semantic loss for reports $X^n$ and semantic weights $u$} if  
$$\ell(X(1);Y|u) \leq \delta $$
for $Y\sim s_{X^n}(y).$

\end{define}

\subsection{Universal summarization}

In the universal setting, the summarizer is no longer aware of the distribution $p\in \mcf{P}(\mcf{X})$ by which $X^n \sim \prod_{j=1}^n p(x(j)),$ 
Since we still assume the end user is aware of this distribution, the summary interpretation remains unchanged.
But, as our results demonstrate, knowing the summary interpretation is of vital importance to cultivating good summarizers. 
Since the summary interpretation is no longer known, the summarizer must be able to adapt itself based upon the report history. 

To measure the performance in this case, we will consider the semantic loss when averaged uniformly over all possible distributions of $P(\mcf{X})$.
In that way, we can ensure that the number of distributions for which the summarizer performs poorly are relatively small. 
\begin{define}
A summarizer $s \in \mcf{P}(\mcf{Y}|\mcf{X}^n)$ has $\delta$-\emph{uniform average semantic loss for semantic weights $u$} if 
$$\int_{\mcf{P}(\mcf{X})} \ell(X(1);Y|u) \dx r \leq \delta $$
for $Y\sim s_{X^n}(y),$ $X^n \sim \prod_{j=1}^n r(x(j))$ and $r$ uniform over $\mcf{P}(\mcf{X}).$ 
\end{define}

\section{Results}

\iftoggle{arxiv}{}{Due to space considerations, only proof sketches are provided here.
For complete proofs see~\textcolor{red}{ADD REFERENCE}}.

Our objective is to find optimal, or close to optimal, summarization algorithms. 
To this end, we first classify what the semantic loss is for a summarizer, and then use that value to determine which summarizer produces the smallest value.
\begin{lemma}\label{lem:3}
Summarizer $s \in \mcf{P}(\mcf{Y}|\mcf{X}^{n})$ has a semantic loss for reports $X^n\sim \prod_{m=1}^n p(x(m))$ and semantic weights $u$ of
\begin{align}
\sum_{\substack{x \in \mcf{X}, \\y \in \mcf{Y}} } p(x) s_{x}(y) \sum_{\mcf{W}\subseteq \mcf{X}} u(x,\mcf{W})(1 - i_{y}(\mcf{W})), \notag 
\end{align}
where $s_{x}(y) = \sum_{x^n: x(1) = x} \left( \prod_{m=2}^n p(x(m)) \right) s_{x^n}(y)$.
\end{lemma}
\begin{cor} \label{cor:3}
The minimum semantic loss for reports $X^n \sim \prod_{m=1}^n p(x(m))$ and semantic weights $u$ is 
$$ \sum_{x} p(x)  \min_{y \in \mcf{Y}} \sum_{\mcf{W}\subseteq\mcf{X}} u(x,\mcf{W})(1 - i_{y}(\mcf{W})).  $$
\end{cor}
\noindent \iftoggle{arxiv}{See Appendix~\ref{app:lem:3} for proof.}{
\begin{proofsketch}
To prove the lemma, use first that the variational distance is
$$\max_{\mcf{\hat X}\subseteq \mcf{X}} q_{X}(\mcf{\hat X}) - i_{Y}(\mcf{\hat X})  \geq q_{X}(\mcf{W}) - i_{Y}(\mcf{W})$$
since $q \in \mcf{P}(\mcf{W}|\mcf{X}).$
Then choosing a distribution $q$ such that $q_{X}(x) \geq i_{Y}(x)$ for all $x \in \mcf{W}$, it follows that $$\sum_{x\in \mcf{X}} \frac{|q_{X}(x) - i_{Y}(x)|}{2} \leq 1 - i_{Y}(\mcf{W}). $$

Setting $s_{x}(y) = 1$ if and only if $y$ minimizes $u(x(1),\mcf{W})(1 - i_{y}(\mcf{W})),$ proves the corollary, since $X(1)\sim p.$
\end{proofsketch}}

Lemma~\ref{lem:3} demonstrates that the semantic loss is the weighted average of the summary interpretation's concentration outside $\mcf{W}$; that is the semantic loss is the weighted average of the various semantic meanings being false under the summary interpretation.
Corollary~\ref{cor:3} also suggests a summarization algorithm to achieve it.
In particular, given reports $x^n$, the summarizer selects the summary $y$ which minimizes
$$\sum_{\mcf{W}} u(x(1),\mcf{W}) \left(1 - \frac{p(\mcf{W}\cap \mcf{X}(y)}{p(\mcf{X}(y))} \right) . $$
To do so though, requires the summarizer know $p$ a priori. 

When moving to the universal setting the value of $p$ is unknown, and instead the distribution $p$ has to be inferred from the reports.  
Here we seek to derive the uniform average semantic loss, for semantic weights $u$, and then find the summarization algorithm to optimize it.
\begin{theorem}\label{thm:4}
Summarizer $s \in \mcf{P}(\mcf{Y}|\mcf{X}^{n})$ has a uniform average semantic loss for semantic weights $u$ of
\begin{align}
& \sum_{\substack{ x^n \in \mcf{X}^{n}  ,\\ y \in \mcf{Y} , \\  \mcf{W}\subseteq \mcf{X}} }   
\frac{s_{x^n}(y) u(x(1),\mcf{W}) }{ |\mcf{T}^n_{(x^n)}| |\mcf{P}_{n}(\mcf{X})| }  \left( 1 - \eta_{x^n,y,\mcf{W}} \right) \notag
\end{align}
where
\begin{align}
\eta_{x^n,y,\mcf{W}} &\defn q^{(x^n)}(\mcf{W}\cap \mcf{X}(y))\notag \\
%& \cdot\! \zeta\left( n+|\mcf{X}|\! -\! (n+|\mcf{X}|+1)\hat q^{(x^n)}(\mcf{X}(y)) , n+|\mcf{X}|\right) ,\notag\\
&\hspace{-35pt} \cdot \sum_{k=0}^{\infty} \frac{\left( n+|\mcf{X}|\! -\! (n+|\mcf{X}|+1)\hat q^{(x^n)}\left(\mcf{X}(y)\right) + k \right)! \left(n+|\mcf{X}|\right)}{\left( n+|\mcf{X}|\! -\! (n+|\mcf{X}|+1)\hat q^{(x^n)}(\mcf{X}(y)) \right)! \left(n+|\mcf{X}|+k\right)!} ,\notag\\
%\zeta(a,b) &\defn \sum_{k=0}^{\infty} \frac{(a+k)! b!}{a!(b+k)!} ~~\forall a,b \in \mathbb{R} \notag, \\
q^{(x^n)}(a) &\defn \frac{n\pi^{(x^n)}(a) + 1}{n+|\mcf{X}|}~~ \forall a \in \mcf{X}, \notag \\
\hat q^{(x^n)}(a) &\defn \begin{cases} \frac{ n\pi^{(x^n)}(a) + 2}{n+|\mcf{X}|+1} &\text{ if } a = x(1) \\ \frac{ n\pi^{(x^n)}(a) +1}{n+|\mcf{X}|+1} &\text{ else} \end{cases}~~ \forall a \in \mcf{X} \notag.
\end{align}
\end{theorem}
\iftoggle{arxiv}{\noindent See Appendix~\ref{app:mp} for proof.}{
\begin{proofsketch}
Writing out explicitly the definition for uniform average semantic loss, it becomes apparent that this essentially relates to the somewhat basic calculus problem of determining 
\begin{align}
\int_{\mcf{P}(\mcf{X})} r(p) \left( \prod_{m=1}^n p(x(m)) \right) \left( 1 - \frac{\sum_{x \in \mcf{X}(y) \cap \mcf{W}} p(x) }{\sum_{x \in \mcf{X}(y)} p(x)} \right) \dx r \label{eq:4:sketch1}
\end{align}
where $r$ is uniform over $\mcf{P}(\mcf{X})$, for an arbitrary $X^n = x^n. $
For each sequence $x^n,$ Equation~\eqref{eq:4:sketch1} can be written as the difference of  integrals
\begin{align}
 &\int_{\mcf{P}(\mcf{X})} r(p^n) \left( \prod_{m=1}^{|\mcf{X}|} p_m^{n \pi^{(x^n)}(m)} \right) \dx r \notag \\
 &\hspace{10pt}-   \int_{\mcf{P}(\mcf{X})} r(p^n) \left( \prod_{m=1}^{|\mcf{X}|} p_m^{n \pi^{(x^n)}(m)} \right)  \frac{\sum_{m = 1}^{|\mcf{X}(y)\cap \mcf{W}|} p_m }{\sum_{m=1}^{ |\mcf{X}(y)|} p_m }  \dx r \label{eq:4:sketch2},
\end{align}
where without loss of generality we have assumed that $\mcf{X} = \{ 1,\dots, |\mcf{X}|\},$ $\mcf{X}(y) = \{ 1 ,\dots, |\mcf{X}(y)|\}$ and $\mcf{X}(y)\cap \mcf{W} = \{ 1 ,\dots, |\mcf{X}(y) \cap \mcf{W}|\}$.

Both integrals are evaluated in the same manner, so focusing on the latter.
The major difficulty in evaluating the intergral is due to the $\frac{1}{\sum_{x =1}^{ |\mcf{X}(y)|} p_m}$ term, which is addressed by using the Taylor series expansion\footnote{In the context of Equation~\eqref{eq:4:sketch2}, when the value for $t\rightarrow 0$, the integrand converges to $0$. This justifies continued use of the representation when $t=0$} of $1/1-t$ yielding 
\begin{align}\frac{1}{\sum_{m =1}^{|\mcf{X}(y)|} p_m } =  \sum_{ k=0}^ \infty \left( \left[ 1 - \sum_{m=1}^{|\mcf{X}(y)|-1} p_m\right] - p_{|\mcf{X}(y)|} \right). \end{align} 

The integral can now be calculated recursively, using 
\begin{align}
    \int_{0}^{c} (c -z)^a z^b ~~\dx z = \frac{a! b!}{(a+b+1)!}  c^{a+b+1} 
\end{align}
from~\cite[Chapter~13,~Section~2]{CT}, and by
recognizing that $r(p^n) = (|\mcf{X}|-1)!$ where the domain $\mcf{P}(\mcf{X})$ is defined recursively by 
$$ p_m \in \left[0, 1 - \sum_{k=1}^{m-1}p_k\right] ~~~\forall m \in \{1,\dots, n-1\}$$
and $p_n = 1 - \sum_{k=1}^{n-1} p_k.$

\end{proofsketch}}

Theorem~\ref{thm:4} though, unlike Lemma~\ref{lem:3}, is not a closed form solution. 
In order to assuage this malady the following approximation is provided.
\begin{lemma}\label{lem:thistooktoolong}
For positive integers $b,~c$ such that $1\leq b<b +2\leq  c$, 
\begin{align}
 \frac{c+1}{c-b}  \leq     \sum_{t = 0}^\infty  \frac{(b+t)!c!}{(c+t)! b!} \leq   \frac{c+1}{c-b} \left( 1 + \varepsilon (c-b)  \right) \notag
%\label{eq:tttl:goal}
\end{align}
where
$$\varepsilon(a) = 3\frac{1 + \ln(a)}{a}  + 4 e^{\frac{1}{12}} \cdot 2^{-\frac{a}{2}}.$$
\end{lemma}

%$2^{-(c - b) + \frac{1}{2} (\log c - \log b)} = 2^{-(c-b) \left( 1 - \frac{1}{2} \frac{\log c - \log b}{c-b} \right)} \geq 2^{-(c-b) \left( 1 -\frac{1}{2b} \right)} \leq 2^{-\frac{c-b}{2}} $
\iftoggle{arxiv}{\noindent See Appendix~\ref{app:tttl} for proof.}{
\begin{proofsketch}
Letting $s(t) = \frac{(b+t)!c!}{(c+t)!b!}$.
The lower bound follows from 
\begin{align}
s(t) = \frac{(b+1) \dots (b+t)}{(c+1) \dots (c+t)} \geq \left( \frac{b+1}{c+1} \right)^t, \label{eq:tttl:lb1}
\end{align}
being plugged into the summation and then observing the result is a geometric series.

The upper bound is a bit more involved. 
In general the summation is split into three different regions depending on the best approximation for the summation.
For earlier terms, the summation terms are similar to a geometric series, while latter terms closely resemble a power series.
A detailed analysis yields
\begin{align}
\sum_{t=0}^{3c \frac{\ln (c-b)}{c-b}} s(t) &\leq \sum_{t=0}^{\infty} \left(\frac{b+3c \frac{\ln (c-b)}{c-b}}{c+3c \frac{\ln (c-b)}{c-b}} \right)^t \\
&=  \frac{c+1}{c-b} \left( 1 + \frac{\ln(c-b)+1}{c-b} \right) , 
\end{align}
similarly 
\begin{align}
\sum_{t=3c \frac{\ln (c-b)}{c-b}}^{2 c}  s(t) &\leq \sum_{t=3c \frac{\ln (c-b)}{c-b}}^{\infty} \left(\frac{b+2c }{c+2c } \right)^t \leq   \frac{c+1}{c-b} \left( \frac{3}{c-b} \right) , 
\end{align}
and 
\begin{align}
\sum_{t=2 c+1}^{\infty}  s(t) &\leq e^{\frac{1}{12b}} \sqrt{\frac{c}{b}} c^{c-b} \sum_{t=2c+1} t^{-(c-b)} \\
&\leq \frac{c+1}{c-b}( 4 e^{\frac{1}{12}} 2^{-\frac{c-b}{2}} ) . 
\end{align}

 \end{proofsketch}
}

Using Theorem~\ref{thm:4} and Lemma~\ref{lem:thistooktoolong} will allow us to construct a theorem analogous to Corollary~\ref{cor:3}.
\begin{theorem}\label{thm:Dear math, Please stop being annoying. Love, Eric}
The minimum uniform average semantic loss for semantic weights $u$, is equal to
\begin{align}
& -\lambda_n + \sum_{\substack{ x^n \in \mcf{X}^{n}  } }   
\frac{ \mu(x^n) }{ |\mcf{T}^n_{(x^n)}| |\mcf{P}_n(\mcf{X})| } \notag
\end{align}
where 
$$\mu(x^n) \defn  \min_{y \in \mcf{Y}} \sum_{\mcf{W}\subseteq \mcf{X}}  u(x(1),\mcf{W}) \left( 1 - \frac{q^{(x^n)}(\mcf{W}\cap \mcf{X}(y))}{\hat q^{(x^n)}( \mcf{X}(y)) } \right)$$
and $q^{(x^n)}$ and $\hat q^{(x^n)}$ from Theorem~\ref{thm:4}, while $\lambda_n$ satisfies 
\begin{align}
&0 < \lambda_n <  u^{*} \left[ \frac{(|\mcf{X}|-1)|\mcf{X}|}{n - \sqrt{n} + |\mcf{X}|-1} \varepsilon(|\mcf{X}|2^{-j}) + \varepsilon(\sqrt{n}) \right] \notag
\end{align}
with $u^{*} = \sum_{x \in \mcf{X}} |\mcf{X}|^{-1} \sum_{\mcf{W} \subseteq \mcf{X}} u(x,\mcf{W}), $ and $\varepsilon$ from Lemma~\ref{lem:thistooktoolong}.

\end{theorem}
\iftoggle{arxiv}{\noindent See Appendix~\ref{app:Dear math, Please stop being annoying. Love, Eric} for proof.}{
\begin{proofsketch}
This result mainly follows from Theorem~\ref{thm:4} and~Lemma~\ref{lem:thistooktoolong}, which together show
\begin{align}
&\eta_{x^n,y,\mcf{W}} \geq  \frac{q^{(x^n)}(\mcf{W}\cap \mcf{X}(y))}{\hat q^{(x^n)}( \mcf{X}(y)) } 
\end{align}
and
\begin{align}
&\eta_{x^n,y,\mcf{W}} \notag\\
&\hspace{5pt}\leq  \frac{q^{(x^n)}(\mcf{W}\cap \mcf{X}(y))}{\hat q^{(x^n)}( \mcf{X}(y)) } \left[ 1 + \varepsilon\left((n+|\mcf{X}|+1)\hat q^{(x^n)}(\mcf{X}(y))\right) \right].
\end{align}
The remainder of the proof is an overly technical analysis showing that the proportion of sequences $(n+|\mcf{X}|+1)\hat q^{(x^n)}(\mcf{X}(y)) < \sqrt{n}$ is less than $$\frac{(|\mcf{X}|-1)|\mcf{X}|}{n - \sqrt{n} + |\mcf{X}|-1},$$
while their error term is at most $\varepsilon(|\mcf{X}|2^{-j})$, where $j$ is the length of the summarizer.
\end{proofsketch}
}

Note that $\lim_{n \rightarrow \infty} \lambda_n = 0$, and thus 
Theorem~\ref{thm:Dear math, Please stop being annoying. Love, Eric} shows that a summarizer that $y$ to minimize 
$$\sum_{\mcf{W}\subseteq \mcf{X}}  u(x(1),\mcf{W}) \left( 1 - \frac{q^{(x^n)}(\mcf{W}\cap \mcf{X}(y))}{\hat q^{(x^n)}( \mcf{X}(y)) } \right)$$
will asymptotically with report history minimize the uniform average semantic loss for semantic weights $u$.
Hence regardless of the set function to characterize semantic meaning, the optimal summarizer still treats the underlying summary interpretation as $\frac{q^{(x^n)}}{\hat q^{(x^n)}}.$
%Furthermore it is interesting to note the similarities between this result and traditional universal source coding results (see~\cite[Chapter~13]{CT} or~\cite[Chapter~6]{csiszar2004information}).
%In both cases we have coding schemes which essentially apply their basic coding scheme to sequence where a distribution close to, but not equal to, the empirical distribution is treated as the true distribution governing the system.

\section{Conclusion}

Going forward it will be important to derive representations for the semantic weights which are practical and perform well in practice.
Indeed, one aspect not previously mentioned is that for any ``optimal'' summary, regardless of how optimal is defined, there are a set of semantic weights such that it is also the optimal summary in our model. 
To see this, consider an optimal (deterministic) summarizer defined by the mapping $x \mapsto y_{x}$, and recognize that this is also an optimal summary in our model for semantic weights 
$$u (x,\mcf{W}) = \begin{cases} 1 &\text{ if } \mcf{W} = \mcf{X}(y_{x}) \\ 0 &\text{otherwise} \end{cases} .$$
While the above is clearly not edifying, it does demonstrate the generality of our model.
Nevertheless, determination of simple semantic weights that perform well in practice would validate the presented model.

\bibliographystyle{ieeetr} \bibliography{this_summ,this}
\iftoggle{arxiv}{\appendices}{}
\iftoggle{arxiv}{

\section{Lemmas}
\begin{lemma}\label{lem:proof:ok}
$$\int_{0}^{c} (c -x)^a x^b ~~\dx x = \frac{a! b!}{(a+b+1)!}  c^{a+b+1} $$
for all non negative integers $a,b$ and real number $c \in [0,1]$.
\end{lemma}

\begin{IEEEproof}

First observe that 
\begin{align}
\int_{0}^c (c-x)^a x^b \dx x &= c^{a+b} \int_{0}^{c} \left( 1 - \frac{x}{c} \right)^a \left(\frac{x}{a} \right)^b \notag \\
&= c^{a+b+1} \int_{0}^1 (1-t)^a t^b \dx t. 
\end{align}
The final integral can be found in \cite{CT}, but we include it for completeness.
Specifically,
\begin{align*}
&\int_{0}^1 (1-t)^a t^b \dx t \\
&= \left[ \frac{1}{b+1}(1-t)^{a}t^{b+1}\right]_{0}^{1} + \frac{a}{b+1} \int_{0}^1 (1-t)^{a-1} t^{b+1} \dx t \\
&=  \frac{a}{b+1} \int_{0}^1 (1-t)^{a-1} t^{b+1} \dx t 
\end{align*}
by using integration by parts (setting $u = (1-t)^a$ and $\dx v = t^b \dx t$). Thus
$$\int_{0}^1 (1-t)^a t^b \dx t = \frac{b!a!}{(a+b+1)!}, $$ from recursion.
\end{IEEEproof}

\begin{cor} \label{cor:proof:ok}
Distribution $r\in \mcf{P}( \mcf{P}(\mcf{X}))$, where $r(p) =(|\mcf{X}|-1)!$, is the uniform distribution over $\mcf{P}(\mcf{X})$.
\end{cor}
\begin{IEEEproof}
First note that $\mcf{P}(\mcf{X})$ is a convex set
$$\set{ (p_1,\dots,p_{|\mcf{X}|}) \in \mathbb{R}^{|\mcf{X}|} : \left( \begin{array}{l}  p_i \in \left[ 0, \bar p_i \right] \\ ~~~~i \in \set{1,\dots,|\mcf{X}|-1},\\ p_{|\mcf{X}|} =  \bar p_{|\mcf{X}|}  \end{array} \right) } ,$$
where $\bar p_i \defn 1- \sum_{j=1}^{i-1} p_j.$
Hence if $r$ is the uniform probability density function over $\mcf{P}(\mcf{X})$ then there exists a constant positive real number $c$ such that $r(p) = c$ and
\begin{align}
\int_{0}^{1} \int_{0}^{\bar p_2} \dots \int_{0}^{\bar p_{|\mcf{X}|-1}} c \dx p_1 \dots \dx p_{|\mcf{X}|-1} = 1.  \label{eq:cor:proof:ok:1}
\end{align}
Using Lemma~\ref{lem:proof:ok} to repeatedly evaluate the LHS of~\eqref{eq:cor:proof:ok:1} yields
$$\int_{0}^{1} \int_{0}^{\bar p_2} \dots \int_{0}^{\bar p_{|\mcf{X}|-1}} c \dx p_1 \dots \dx p_{|\mcf{X}-1|} = \frac{c}{(|\mcf{X}|-1)!} ,$$
hence 
$$r(p) = c =(|\mcf{X}|-1)!.$$

\end{IEEEproof}

\begin{lemma}\label{lem:approx:1}
$$\frac{(k+1)^{-(t-1)}}{t-1} \leq \sum_{j= k +1 }^\infty j^{-t} \leq \frac{k^{-(t-1)}}{t-1} $$
for all positive integers $k$ and real numbers $t>1$.
\end{lemma}

\iftoggle{arxiv}{\begin{IEEEproof}
First note that $k^{-t}$ is convex in $k$ since $t>1$, and thus
\begin{align*}
\frac{(k+1)^{-t+1}}{t-1} &=  \int_{k+1}^\infty x^{-t} \dx x \leq \sum_{j= k +1 }^\infty j^{-t} \\
&\hspace{20pt}\leq \int_{k+1}^\infty (x-1)^{-t} \dx x = \frac{k^{-t+1}}{t-1}.
\end{align*}
\end{IEEEproof}}{}

%\begin{lemma}\label{lem:approx:2}
%$$ \sum_{j = k}^\infty x^j = - x^k \frac{1}{1-x} ,$$
%for all positive real numbers $x \in [0,1).$
%\end{lemma}
%\begin{IEEEproof}
%The lemma follows primarily by the Taylor series approximation of $1/(1-x),$ in specific
%\begin{align}
%\sum_{j = k}^\infty x^j&=  x^{k} \sum_{j = 0}^\infty x^j  = x^k \frac{1}{1-x} ,    \label{eq:ln approx:1}
%\end{align}
%for all $x \in [0,1).$ 
%\end{IEEEproof}

\begin{lemma}\label{lem:tttl_support1}
For $a > b$ positive real numbers,
$$ \left( \frac{a}{b} \right)^b e^{-(a-b)} \leq 1 .$$
\end{lemma}
\iftoggle{arxiv}{\begin{IEEEproof}
For any positive real number $y$, by definition 
\begin{align}
\lim_{x \rightarrow \infty} \left( 1 + \frac{y}{x} \right)^{x} = e^{y}.
\end{align}
Hence we must show that $f(x,y) \defn \left( 1 + \frac{y}{x} \right)^{x} $ is a monotonically increasing function of $x$ for all $x \geq 1$, since then $$\left( \frac{a}{b} \right)^b =  \left( 1 +  \frac{a-b}{b} \right)^b  \leq e^{a-b}.$$ Clearly though 
\begin{align}
\frac{\partial}{\partial x} f(x,y) = \left( 1 + \frac{y}{x} \right)^{x} \left[ \ln \left( \frac{x+y}{x} \right) + \frac{x}{y+x} - 1 \right]. 
\end{align}
This derivative is always positive, since the function $g(u) \defn \ln u + \frac{1}{u} - 1 \geq 0$ for all $u \geq 1$. Indeed
$g(u) \geq g(1) $ since 
$$\frac{\dx}{\dx u} g(u) = \frac{u-1}{u^2} \geq 0 ~\forall u \geq 1,$$
and $g(1) = \ln 1 + 1 -1 = 0.$
\end{IEEEproof}}{}

\begin{lemma}\label{lem:tttl_support2}
For positive integers $a > b$, and positive integer $j$,  $$\frac{e^{-b} \left( 1 +\frac{b}{j} \right)^{b+j + \frac{1}{2}}}{e^{-a} \left( 1 +\frac{a}{j} \right)^{a+j + \frac{1}{2}}} \leq 1.$$
\end{lemma}
\iftoggle{arxiv}{\begin{IEEEproof}

That $ \frac{f(b)}{f(a)} \leq 1$, where $f(x) \defn e^{-x} \left( 1 +\frac{x}{j} \right)^{x+j + \frac{1}{2}}$ follows directly from
$$ \frac{\partial }{\partial x} f(x) = f(x) \left[ \frac{1}{2(x+j)} + \ln\left( 1 + \frac{x}{j} \right) \right] \geq 0$$
for all positive values of $x$.
\end{IEEEproof}}{}

\begin{lemma}\label{lem:super loose}
For positive integers $b,c$ such that $1 \leq  b < b+2 \leq c ,$
$$\sqrt{\frac{c}{b}} \leq 2^{\frac{c-b}{2}}.$$
\end{lemma}
\iftoggle{arxiv}{\begin{IEEEproof}

For $b =1 $ the lemma follows because $ c \geq b+2 \geq 3$ and $c \leq 2^{c-1} $ for all $ c \geq 2.$

For $b \geq 2,$ 
$$\logt c - \logt b \leq (c-b) \cdot \max_{x \in [b , c]} \frac{\dx \logt x}{\dx x} =  \frac{c-b}{b \ln 2} < c-b $$
implies that
$$\sqrt{\frac{c}{b}} = 2^{\frac{1}{2}\left[ \logt c - \logt b\right]} \leq 2^{\frac{c-b}{2}} .$$
\end{IEEEproof}}{}

}{}
\iftoggle{arxiv}{\section{Lemma~\ref{lem:3}}\label{app:lem:3}

\begin{IEEEproof}
If 
\begin{align}
\inf_{q \in \mcf{P}(\mcf{W}|\mcf{X})}  \sum_{x \in \mcf{X}} \frac{1}{2} | q_{\hat x}(x) - i_{y}(x) | = 1 - i_{y}(\mcf{W}), \label{eq:3:key_equation}
\end{align}
then clearly 
\begin{align} &\ell(X;Y|u) &= \sum_{\substack{x \in \mcf{X}, \\ y \in \mcf{Y}}}  p(x) s_{x}(y)  \sum_{\mcf{W} \subseteq \mcf{X}} u(x,\mcf{W}) ( 1- i_{y}(\mcf{W}))  
\end{align}
by definition.

To prove Equation~\eqref{eq:3:key_equation}, first obtain a lower bound to the LHS of Equation~\eqref{eq:3:key_equation} via
\begin{align}
 \sum_{x \in \mcf{X}} \frac{1}{2} | q_{\hat x}(x) - i_{y}(x) | &= \max_{\mcf{\tilde X} \subseteq \mcf{X}}  q_{x}(\mcf{\tilde X}) -i_{y}(\mcf{\tilde X}) \label{eq:3:key_equation:lb1}\\
 &\geq q_{x}(\mcf{W}) - i_{y}(\mcf{W}) = 1 - i_{y}(\mcf{W}),
\end{align}
where Equation~\eqref{eq:3:key_equation:lb1} is an alternative equation for the variational distance. 
Next let $\tilde q \in \mcf{P}(\mcf{W}|\mcf{X})$ be any distribution such that $q_{\hat x} (a)\geq i_{y}(a) $ for all $a \in \mcf{W}.$
Now obtain an upper bound to Equation~\eqref{eq:3:key_equation} via
\begin{align}
&\hspace{-20pt} \inf_{q \in \mcf{P}(\mcf{W}|\mcf{X})}  \sum_{x \in \mcf{X}} \frac{1}{2} | q_{\hat x}(x) - i_{y}(x) | \notag \\
&\leq  \sum_{x \in \mcf{X}} \frac{1}{2} | \tilde q_{\hat x} (x) - i_{y}(x) | \\
&= \frac{\sum_{a \in \mcf{W}} \tilde q_{\hat x}(x) - i_{y}(x) }{2} + \frac{\sum_{a \in \mcf{X} -\mcf{W}}  i_{y}(x) }{2} \\
&= \frac{1 - i_{y}(\mcf{W}) }{2} + \frac{ i_{y}(\mcf{X} - \mcf{W}) }{2}  = 1 - i_{y}(\mcf{W}).
\end{align}

\end{IEEEproof}}{}
\iftoggle{arxiv}{\section{Proofs of main results} \label{app:mp}

\begin{IEEEproof}

To begin the proof, note the uniform average semantic loss for a given summarizer can be written
\begin{align}
&  \sum_{\substack{ x^n \in \mcf{X}^{n}  ,\\ y \in \mcf{Y} , \\  \mcf{W}\subseteq \mcf{X}} } s_{x^n}(y) u(x(1),\mcf{W})  \alpha(\mcf{W},x^n,y)
%&   \sum_{\substack{ x^n \in \mcf{X}^{n}  ,\\ y \in \mcf{Y} , \\  \mcf{W}\subseteq \mcf{X}} } u(x,\mcf{W}) s_{x^n}(y)  
%\notag \\
% &\int_{\mcf{P}(\mcf{X})} \left( \prod_{\tilde x \in \mcf{X} } p_{\tilde x}^{n \pi^{(x^n)}(\tilde x)} \right) r(p) \left( 1 - \frac{p(\mcf{W}\cap \mcf{X}(y)}{p(\mcf{X}(y)} \right) \dx p^{|\mcf{X}|} 
\label{eq:proof:KL|PU}
\end{align}
where 
$$\alpha(\mcf{W},x^n,y) \hspace{-2pt} = \hspace{-5pt} \int_{\mcf{P}(\mcf{X})} \hspace{-15pt} r(p) \left( \prod_{m = 1}^n  p(x(m))  \hspace{-4pt}\right)  \hspace{-3pt} \left(  \hspace{-2pt}1 -  \frac{p(\mcf{W}\cap \mcf{X}(y)}{p(\mcf{X}(y))} \right)  \dx r , $$
and $r$ is uniform over $\mcf{P}(\mcf{X})$, due to the integral function being linear.
The proof proceeds by evaluating $\alpha(\mcf{W},x^n,y),$ and specifically showing
\begin{align}
\alpha(\mcf{W},x^n,y) &=\frac{1}{ |\mcf{T}^n_{(x^n)}| |\mcf{P}_{n}(\mcf{X})| }  \left( 1 - \eta_{x^n,y,\mcf{W}} \right) . \label{eq:proof:lazy0}
\end{align}

To help in evaluating the integrals, assume that $\mcf{X} =\{ 1,\dots, |\mcf{X}|\}$, and let
\begin{align*}
p_m &\defn p(m) ,\\
\bar p_m &\defn 1 - \sum_{m=1}^{k-1} p_k ,\\
t_m &\defn  n\pi^{( x^n)}(m), \\
\bar t_m &\defn |\mcf{X}| - m  + \sum_{k=m}^{|\mcf{X}|} t_k = |\mcf{X}|-m + n - \sum_{k=1}^{m-1} t_k ,
\end{align*}
for all $m \in \mcf{X}.$
Of importance throughout the proof will be that 
\begin{align}
\bar p_m = \bar p_{m-1} - p_{m-1} \label{eq:proof:helpful}
\end{align}
for all integers $m \in \mcf{X},$ and that
\begin{align}
\bar t_m = \bar t_{m+1} + t_m +1 \label{eq:proof:helpful2}.
\end{align}
Two notable values are
$\bar t_1 = n + |\mcf{X}|-1$ and $\bar p_1 = 1$.
Also, without loss of generality assume that $\mcf{W}\cap \mcf{X}(y) = \{1,\dots, w\}$, $\mcf{X}(y) = \{1,\dots,z\}$ and $x(1) = 1.$

With this new notation
\begin{align}
\alpha(\mcf{W},x^n,y) = &(|\mcf{X}|-1)! \int_{\mcf{P}(\mcf{X})}  \left( \prod_{m =1}^{|\mcf{X}|} p_m^{t_m} \right) \dx p^n \notag \\
& \hspace{-15pt} - (|\mcf{X}|-1)! \int_{\mcf{P}(\mcf{X})}  \left( \prod_{m =1}^{|\mcf{X}|} p_m^{t_m} \right) \frac{\sum_{m =1}^{w} p_m}{\sum_{m=1}^z p_m}  \dx p^n \label{eq:proof:lazy1}
\end{align}
where $\dx p^n = \dx p_n \dx p_{n-1} \dots \dx p_1$, since $r(p) = (|\mcf{X}|-1)!$ by corollary~\ref{cor:proof:ok} and $\mcf{P}(\mcf{X})$ is the convex set
$$\set{ (p_1,\dots,p_{|\mcf{X}|}) \in \mathbb{R}^{|\mcf{X}|} \hspace{-3pt} : \hspace{-3pt}\left( \begin{array}{l}  p_m \in \left[ 0, \bar p_m \right] \\ ~~~~m \in \set{1,\dots,|\mcf{X}|-1},\\ p_{|\mcf{X}|} =  \bar p_{|\mcf{X}|}  \end{array} \right) \hspace{-3pt}} .$$
Of the two integrals in~\eqref{eq:proof:lazy1} we shall only show
\begin{align}
\int_{\mcf{P}(\mcf{X})} \hspace{-3pt} \left( \prod_{m =1}^{|\mcf{X}|} p_m^{t_m} \right) \hspace{-2pt} \frac{\sum_{m =1}^{w} p_m}{\sum_{m=1}^z p_m} \hspace{-2pt} \dx p^n \hspace{-3pt} = \hspace{-3pt} \frac{\eta_{x^n,y,\mcf{W}}}{(|\mcf{X}|-1)! |\mcf{T}_{(x^n)}^n||\mcf{P}_n(\mcf{X})|}     \label{eq:proof:lazy2}
\end{align}
since 
\begin{align}
\int_{\mcf{P}(\mcf{X})}  \left( \prod_{m =1}^{|\mcf{X}|} p_m^{t_m} \right)  \dx p^n = \frac{1}{(|\mcf{X}|-1)! |\mcf{T}_{(x^n)}^n||\mcf{P}_n(\mcf{X})|}     \label{eq:proof:lazy3}
\end{align}
follows similarly. 
At this point note that Equation~\eqref{eq:proof:lazy0} directly follows from Equations~\eqref{eq:proof:lazy1},~\eqref{eq:proof:lazy2} and~\eqref{eq:proof:lazy3}, so validating Equation~\eqref{eq:proof:lazy3} would finish the proof.
We shall prove Equation~\eqref{eq:proof:lazy2} through a rather tedious recursion process.
To aid in this recursion we shall, in an abuse of notation, write
$\mcf{\tilde P}(k)$ to denote the convex set
$$\set{ (p_1,\dots,p_{k}) \in \mathbb{R}^{k} : \left( \begin{array}{l}  p_m \in \left[ 0, \bar p_m \right] \\ ~~~~m \in \set{1,\dots,k-1}  \end{array} \right) } ,$$
and use $\dx p^{k} $ to denote the differential sequence $\dx p_k \dx p_{k-1}\dots \dx p_1.$

Write the LHS of~\eqref{eq:proof:lazy2} 
\begin{align}
& \int_{\mcf{\tilde P}(|\mcf{X}|-2)}  \left( \prod_{m=1}^{|\mcf{X}|-2} p_m^{t_m} \right)  \frac{\sum_{m=1}^{w} p_m}{\sum_{m=1}^{z} p_m }  \notag \\
& \cdot \left( \int_{0}^{\bar p_{|\mcf{X}|-1}}  p_{|\mcf{X}|-1}^{t_{|\mcf{X}|-1}} \left( \bar p_{|\mcf{X}|-1} - p_{|\mcf{X}|-1}\right)^{t_{|\mcf{X}|}}  \hspace{-3pt} \dx p_{|\mcf{X}|-1} \right) \hspace{-3pt} \dx p^{|\mcf{X}|-2} \label{eq:proof:m-1},
\end{align}
by using $p_{|\mcf{X}|} = \bar p_{|\mcf{X}|} = \bar p_{|\mcf{X}|-1} - p_{|\mcf{X}|-1} $, via Equation~\eqref{eq:proof:helpful}. 
The inner integration can be performed via Lemma~\ref{lem:proof:ok} yielding
\begin{align}
& \frac{t_{|\mcf{X}|}! t_{|\mcf{X}|-1}!}{\bar t_{|\mcf{X}|-1}!}\int_{\mcf{\tilde P}(|\mcf{X}|-2)}  \hspace{-4pt}\left( \hspace{-2pt} \prod_{m=1}^{|\mcf{X}|-2} \hspace{-4pt} p_m^{t_m} \hspace{-4pt} \right) \hspace{-4pt} \frac{\sum_{m=1}^{w} p_m}{\sum_{m=1}^{z} p_m } \bar p_{|\mcf{X}|-1}^{\bar t_{|\mcf{X}|-1}} \dx p^{|\mcf{X}|-2} \notag \\
&= \frac{t_{|\mcf{X}|}! t_{|\mcf{X}|-1}!}{\bar t_{|\mcf{X}|-1}!} \int_{\mcf{\tilde P}(|\mcf{X}|-3)} \left( \prod_{m=1}^{|\mcf{X}|-3} p_m^{t_m} \right) \frac{\sum_{m=1}^w p_m}{\sum_{m=1}^{z} p_m }  \notag \\
& \cdot \left(\int_{0}^{\bar p_{|\mcf{X}|-2}}  \hspace{-3pt}  p_{|\mcf{X}|-2}^{ t_{|\mcf{X}|-2}} \left( \bar p_{|\mcf{X}|-2} - p_{|\mcf{X}|-2}\right)^{\bar t_{|\mcf{X}|-1}}  \hspace{-3pt} \dx p_{|\mcf{X}|-2} \hspace{-3pt} \right) \hspace{-3pt} \dx p^{|\mcf{X}|-3} \label{eq:proof:m-2}
\end{align}
where~\eqref{eq:proof:m-2} follows via Equation~\eqref{eq:proof:helpful}, this time to show $\bar p_{|\mcf{X}|-1} = \bar p_{|\mcf{X}|-2} - p_{|\mcf{X}|-2}$ term.
This process of using Lemma~\ref{lem:proof:ok} to evaluate the integral, and then using Equation~\eqref{eq:proof:helpful} to put the result into a form which can be evaluated using Lemma~\ref{lem:proof:ok} can be repeated to evaluate the integrals over $p_{|\mcf{X}|-2},\dots,p_{z+1}$; doing so yields
\begin{align}
&\frac{ \prod_{m=z+1}^{|\mcf{X}|} t_m! }{\bar t_{z+1} !} \hspace{-1pt} \int_{\mcf{\tilde P}(z)} \hspace{-3pt} \left( \prod_{m=1}^{z} \hspace{-2pt} p_m^{t_m} \hspace{-2pt} \right) \left(\bar p_{z} - p_{z} \right)^{\bar t_{z+1}}  \hspace{-2pt} \frac{\sum_{m=1}^{w}p_m}{\sum_{m=1}^{z} p_m }  \dx p^{z} 
\label{eq:proof:ell}.
\end{align}

At this point the recursion no longer directly applies since the next variable of integration, $p_{z}$, is contained in the denominator of the fraction. 
To address this, use the Taylor series expansion
%\footnote{The numerator is included so that the LHS and RHS both converge to $0$ when $\sum_{m=1}^z p_m \rightarrow 0$. Since both have the same limit point, we can continue to use the function Hence thewhen evaluated in the integral For proof see Appendix~\ref{app:roc}. } 
of $\frac{1}{1 - x}$, specifically as follows
\begin{align} 
\frac{1}{\sum_{m =1}^{z} p_{m}} &=  \frac{1}{1 - \left( \bar p_{z} - p_{z} \right)} =   \left( \bar p_{z} - p_{z} \right)^k \label{eq:proof:taylor}.
\end{align}
Plugging~\eqref{eq:proof:taylor} into~\eqref{eq:proof:ell} and exchanging the summations and integrals results in
\begin{align}
&\frac{ \prod_{m=z+1}^{|\mcf{X}|} t_m! }{\bar t_{z+1} !}   \hspace{-2pt}\sum_{k=0}^{\infty}  \sum_{\hat m =1}^{w} \int_{\mcf{\tilde P}_z} \hspace{-4pt} \left( \prod_{m=1}^{z} p_m^{t_m} \hspace{-2pt} \right) \hspace{-2pt} \left(\bar p_{z} - p_{z} \right)^{\bar t_{z+1} + k } \hspace{-2pt}p_{\hat m} \dx p^{z} 
\label{eq:proof:ellk}.
\end{align}
From here, evaluating all remaining integrals using the recursive process by which Equation~\eqref{eq:proof:m-2} and~\eqref{eq:proof:ell} are obtained yields
\begin{align}
&\sum_{k=0}^\infty \sum_{\hat m = 1}^{w} \frac{ \left( \bar t_{z+1} + k \right)!(t_{\hat m}+1)\prod_{m=1}^{|\mcf{X}|} t_m! }{ \bar t_{z+1}! \left( \bar t_{1} + k + 1 \right)!}
\label{eq:proof:ellk2_diff2}.
\end{align}
Then
\begin{align}
&\frac{1}{(|\mcf{X}|-1)!} \frac{\prod_{m=1}^{|\mcf{X}|} t_m!}{n!} \frac{n!(|\mcf{X}|-1)!}{(n+|\mcf{X}|-1)!}  \notag\\
&\hspace{5pt} \cdot  \left( \sum_{\hat m = 1}^{w} \frac{t_{\hat m}+1}{n+|\mcf{X}|} \right) \left( \sum_{k=0}^\infty   \frac{(n+|\mcf{X}|)! \left( \bar t_{z+1} + k \right)! }{ \bar t_{z+1}! \left( \bar t_{1} + k + 1 \right)!} \right) 
\label{eq:proof:ellk2_diff222222}
\end{align}
follows from ``simplifying''~\eqref{eq:proof:ellk2_diff2}.
Equation~\eqref{eq:proof:lazy2} is therefore verified, completing the proof, having shown the LHS equals~\eqref{eq:proof:ellk2_diff222222} since
\begin{align}
 \frac{\prod_{m=1}^{|\mcf{X}|} t_m!}{n!} &= \frac{\prod_{m=1}^{|\mcf{X}|} (n \pi^{(x^n)}(m))!}{n!} = |\mcf{T}^n_{(x^n)}| \notag \\
\frac{n!(|\mcf{X}|-1)!}{(n+|\mcf{X}|-1)!} &= |\mcf{P}_n(\mcf{X})|\notag \\
 \sum_{\hat m = 1}^{w} \frac{t_{\hat m}+1}{n+|\mcf{X}|} &=  \sum_{\hat m = 1}^{w} \frac{n\pi^{(x^n)}(\hat m)+1}{n+|\mcf{X}|}%\notag \\
 %&= \sum_{\hat m = 1}^{w} q^{(x^n)}(\hat m)  
 = q^{(x^n)}(\mcf{W}\cap \mcf{X}(y))\notag \\
\bar t_1 + k + 1 &= n+|\mcf{X}| \notag \\
\bar t_{z+1}& = n + |\mcf{X}| + k - 1 - \sum_{m=1}^{z} (n \pi^{(x^n)}(m) + 1) \notag \\
&= n+|\mcf{X}| - (n+|\mcf{X}|+1) \hat q^{(x^n)}(\mcf{X}(y)) \notag .
\end{align}

\end{IEEEproof}

\begin{comment}
\section{Radius of convergence} \label{app:roc}

\begin{align}
&\lim_{\sum_{m=1}^{z} p_m \rightarrow 0}\frac{p_1\sum_{m =1}^{w} p_{m}}{\sum_{m =1}^{z} p_{m}} \notag\\
&=  \lim_{\sum_{m=1}^{z} p_m \rightarrow 0}  p_1 \sum_{m =1}^{w} p_{m} \sum_{k=0}^{\infty}  \left( \bar p_{z} - p_{z} \right)^k = 0
\end{align}
allows for the integral 
 \end{comment}}{}
\iftoggle{arxiv}{\section{Proof of Lemma~\ref{lem:thistooktoolong}}\label{app:tttl}

\begin{IEEEproof}
Let $s(j) \defn \frac{(b+j)!c!}{(c+j)!b!}$.

The lower bound follows primarily because 
\begin{align}
s(j) = \frac{(b+1) \dots (b+j)}{(c+1) \dots (c+j)} \geq \left( \frac{b+1}{c+1} \right)^j. \label{eq:tttl:lb1}
\end{align}
Indeed given Equation~\eqref{eq:tttl:lb1}  
\begin{align}
 \sum_{j = 0}^\infty \frac{(b+j)!c!}{(c+j)! b!} \geq  \sum_{j = 0}^\infty  \left( \frac{b+1}{c+1} \right)^j &=  \frac{c+1}{c-b} 
\end{align}
since $\frac{b}{c} < 1$.

The upper bound is a bit more involved. Begin by letting $\phi$ and $\rho$ be arbitrary positive integers the values for which will be specified later.
We are going to split the summation of $s(j)$ into three distinct regions, each of which will be given a different upper bound. 
This is done because $s(j)$ behaves differently depending on where it is in the summation terms, with earlier terms more resembling the Taylor series of $1/(1-x) $ while the later terms more resemble a geometric series. 
In specific,
\begin{align}
\sum_{j=0}^\infty  s(j) &\leq \sum_{j=0}^\phi  \left( \frac{b+\phi}{c+\phi} \right)^j + \sum_{j=\phi+1}^\rho  \left( \frac{b+\rho}{c+\rho} \right)^j \notag \\
&\hspace{-35pt} + \sum_{j=\rho+1}^\infty e^{\frac{1}{12b}} \left( \frac{c^{c+1/2}}{b^{b+1/2}} e^{-(c-b)} \right) j^{b-c}  \left[\frac{e^{-b} \left( 1 + \frac{b}{j} \right)^{b+j+1/2} }{e^{-c}\left( 1 + \frac{c}{j} \right)^{c+j+1/2} } \right] 
\end{align}
since $\frac{b+k}{c+k} = \frac{b}{c} + \frac{k}{c} \left( 1 - \frac{b+k}{c+k} \right) > \frac{b}{c} $ while
\begin{align}
s(j) &\leq e^{\frac{1}{12b}} \left( \frac{c^{c+1/2}}{b^{b+1/2}} e^{-(c-b) } \right) j^{b-c}  \left[\frac{e^{-b} \left( 1 + \frac{b}{j} \right)^{b+j+1/2} }{e^{-c}\left( 1 + \frac{c}{j} \right)^{c+j+1/2} } \right]  \label{eq:tttl:2b}
\end{align}
follows by replacing all factorials with their appropriate counter parts from Robbin's sharpening of Stirling's formula\footnote{$$ \sqrt{2 \pi} n^{n+\frac{1}{2}} e^{-n + \frac{1}{12 n + 1} }  \leq n! \leq \sqrt{2 \pi} n^{n+\frac{1}{2}} e^{-n + \frac{1}{12 n} }  $$
for all positive integers $n$}. 
%Furthermore $\zeta_j \leq \frac{1}{12a} < \zeta^+ \defn \frac{1}{12b}$ %and $\zeta_j \geq - \zeta^+$ 
%for all positive integers $j$ since $a > b$. 
Now extending the first two summations to infinity, and making use of Lemmas~\ref{lem:tttl_support1} and~\ref{lem:tttl_support2} gives 
\begin{align}
\sum_{j=0}^\infty  s(j) &\leq \sum_{j=0}^\infty  \left( \frac{b+\phi}{c+\phi} \right)^j + \sum_{j=\phi+1}^\infty \left( \frac{b+\rho}{c+\rho} \right)^j \notag \\
&\hspace{5pt} + e^{\frac{1}{12b}} \sqrt{\frac{c}{b}} c^{c-b} \sum_{j=\rho+1}^\infty   j^{-(c-b)} \label{eq:tttl:b1}.
\end{align}
Since $\rho$ was arbitrary
\begin{align}
\sum_{j=0}^\infty  s(j) 
%&\leq\frac{c+1}{c-b} \left( 1  + \frac{\phi - 1}{c+1} \right) + \left( \frac{b+2c}{3c} \right)^{\phi+1} \frac{3c}{c-b}\notag \\
%& \hspace{20pt} +   e^{\frac{1}{12b}}  \frac{2c \sqrt{c}}{(c-b-1)\sqrt{b}}   2^{-(c-b)}   \label{eq:tttl:ub1},
&\leq\frac{c+1}{c-b} \left( 1  + \frac{\phi - 1}{c+1} \right) +  e^{- \frac{c-b}{3c} (\phi+1)} \frac{3c}{c-b}\notag \\
& \hspace{20pt} +   e^{\frac{1}{12b}}  \frac{2c \sqrt{c}}{(c-b-1)\sqrt{b}}   2^{-(c-b)}  \notag \\
&\leq \frac{c+1}{c-b} \left( 1  + \frac{\phi }{c} + 3 e^{- \frac{c-b}{3c} \phi}  + 4e^{\frac{1}{12}}   2^{-\frac{c-b}{2}} \right) 
%\notag \\
%& \hspace{20pt} +   e^{\frac{1}{12b}}  \frac{2c \sqrt{c}}{(c-b-1)\sqrt{b}}   2^{-(c-b)}  
\label{eq:tttl:ub1},
\end{align}
follows from Equations~\eqref{eq:tttl:b1} by setting $\rho=2c$ and then recognizing the first two summations as Taylor series expansions of $1/(1-x)$, using~\cite[Lemma~10.5.3]{CT}\footnote{For $0\leq x,y\leq 1, ~n > 0,$ $$(1-xy)^n\leq 1 - x + e^{-yn}.$$} to show
$$\left( \frac{b + 2c}{3c}\right)^{\phi+1} = \left( 1 - \frac{c-b}{3c} \right)^{\phi+1} \leq e^{- \frac{c-b}{3c} (\phi+1)},$$
and for the final summation using Lemmas~\ref{lem:approx:1},~\ref{lem:super loose} and that $c \geq b + 2 > b \geq 1$.

Choosing $\phi = 3c \frac{\ln(c-b)}{c-b} $ in Equation~\eqref{eq:tttl:ub1} yields 
\begin{align}
\sum_{j=0}^\infty s(j) &\leq  \frac{c+1}{c-b} \left( 1 + 3\frac{1 + \ln(c-b)}{c-b}  + 4e^{\frac{1}{12}}  2^{-\frac{c-b}{2}}  \right)
\label{eq:tttl:ub1prime}.
\end{align}

\end{IEEEproof}}{}
\iftoggle{arxiv}{\section{Proof of Theorem~\ref{thm:Dear math, Please stop being annoying. Love, Eric}}\label{app:Dear math, Please stop being annoying. Love, Eric}

\begin{IEEEproof}
First, 
\begin{align}
&\eta_{x^n,y,\mcf{W}} \geq  \frac{q^{(x^n)}(\mcf{W}\cap \mcf{X}(y))}{\hat q^{(x^n)}( \mcf{X}(y)) } 
\end{align}
by Lemma~\ref{lem:thistooktoolong}, implying that 
\begin{align}
& \min_{s \in \mcf{P}(\mcf{Y}|\mcf{X}^n)} \sum_{\substack{ x^n \in \mcf{X}^{n}  ,\\ y \in \mcf{Y} , \\  \mcf{W}\subseteq \mcf{X}} }   
\frac{s_{x^n}(y) u(x(1),\mcf{W}) }{ |\mcf{T}^n_{(x^n)}| |\mcf{P}_{n}(\mcf{X})| }  \left( 1 - \eta_{x^n,y,\mcf{W}} \right) \notag \\
&\leq  \sum_{\substack{ x^n \in \mcf{X}^{n} , \\  \mcf{W}\subseteq \mcf{X}} }   
\frac{ u(x(1),\mcf{W}) }{ |\mcf{T}^n_{(x^n)}| |\mcf{P}_{n}(\mcf{X})|  } \min_{y \in \mcf{Y}} \left( 1 - \frac{q^{(x^n)}(\mcf{W}\cap \mcf{X}(y))}{\hat q^{(x^n)}( \mcf{X}(y)) } \right) . \label{eq:dmplsbale:t1}
\end{align}

Likewise 
\begin{align}
&\eta_{x^n,y,\mcf{W}} \notag\\
&\hspace{10pt}\leq  \frac{q^{(x^n)}(\mcf{W}\cap \mcf{X}(y))}{\hat q^{(x^n)}( \mcf{X}(y)) } \left[ 1 + \varepsilon\left((n+|\mcf{X}|+1)\hat q^{(x^n)}(\mcf{X}(y))\right) \right]
\end{align}
by Lemma~\ref{lem:thistooktoolong}, thus showing (which we will do momentarily)
\begin{align}
& \sum_{\substack{ x^n \in \mcf{X}^{n}  ,\\ y \in \mcf{Y} , \\  \mcf{W}\subseteq \mcf{X}} }  \frac{s_{x^n}(y) u(x(1),\mcf{W}) }{ |\mcf{T}^n_{(x^n)}| |\mcf{P}_{n}(\mcf{X})|  } \varepsilon\left((n+|\mcf{X}|+1)\hat q^{(x^n)}(\mcf{X}(y))\right) \notag \\
 &\hspace{10pt} \leq u^* \left(  \frac{(|\mcf{X}|-1)|\mcf{X}|}{n - \sqrt{n} + |\mcf{X}|-1} \varepsilon(|\mcf{X}|2^{-j}) + \varepsilon(\sqrt{n})   \right) \label{eq:dmplsbale:nts}
\end{align}
implies 
\begin{align}
& \min_{s \in \mcf{P}(\mcf{Y}|\mcf{X}^n)} \sum_{\substack{ x^n \in \mcf{X}^{n}  ,\\ y \in \mcf{Y} , \\  \mcf{W}\subseteq \mcf{X}} }   
\frac{s_{x^n}(y) u(x(1),\mcf{W}) }{ |\mcf{T}^n_{(x^n)}| |\mcf{P}_{n}(\mcf{X})|  }  \left( 1 - \eta_{x^n,y,\mcf{W}} \right) \notag \\
&\geq  u^* \left( \frac{(|\mcf{X}|-1)|\mcf{X}|}{n - \sqrt{n} + |\mcf{X}|-1} \varepsilon(|\mcf{X}|2^{-j}) + \varepsilon(\sqrt{n})   \right) \notag \\
&+\sum_{\substack{ x^n \in \mcf{X}^{n} , \\  \mcf{W}\subseteq \mcf{X}} }   
\frac{ u(x(1),\mcf{W}) }{ |\mcf{T}^n_{(x^n)}| |\mcf{P}_{n}(\mcf{X})|  } \min_{y \in \mcf{Y}} \left( 1 - \frac{q^{(x^n)}(\mcf{W}\cap \mcf{X}(y))}{\hat q^{(x^n)}( \mcf{X}(y)) } \right) \label{eq:dmplsbale:t2}.
\end{align}
Combining Equation~\eqref{eq:dmplsbale:t1} and~\eqref{eq:dmplsbale:t2} proves the theorem. 

Returning to prove Equation~\eqref{eq:dmplsbale:nts}, we will need the three following technical results.

The first technical result is
\begin{align}
\frac{\left| \left\{ x^n \in \mcf{X}^n : \begin{array}{r l} \pi^{(x^n)}&= \rho,\\x(1) &= a  \end{array} \right\}\right|}{|\mcf{T}_{(x^n)}^n|} = \rho(a) \label{eq:dmplsbale:tr1}
\end{align}
for all $\rho \in \mcf{P}_n(\mcf{X})$, $x^n$ such that $\pi^{(x^n)} = \rho,$ and $a \in \mcf{X}.$
Indeed, this is a straightforward combinatorial result since the term in the is simply $|\mcf{T}^{n-1}_{(x(2),x(3),\dots,x(n))}|$, that is 
$$ \left(\!\begin{matrix} n-1 \\ n\rho(1), \dots, n \rho(a-1), n\rho(a) -1 , n\rho(a+1),\dots,n\rho(|\mcf{X}|) \end{matrix} \! \right), $$
because fixing the empirical distribution of $n$-length sequences over $\mcf{X}$, and fixing the first symbol in the sequence, also fixes the empirical distribution of the second through $n$-th symbols.

The second technical result is
\begin{align}
\frac{\left| \{ \rho \in \mcf{P}_{n}(\mcf{X}) :  n\rho (a) < \sqrt{n} \} \right|}{|\mcf{P}_{n}(\mcf{X})|} \leq \frac{(|\mcf{X}|-1)\sqrt{n}}{n-\sqrt{n}+|\mcf{X}|-1}.  \label{eq:dmplsbale:tr2}
\end{align}
This technical result is a consequence of
\begin{align}
\frac{\left| \{ \rho \in \mcf{P}_{n}(\mcf{X}) :  n\rho (a) \geq \sqrt{n} \} \right|}{|\mcf{P}_n(\mcf{X})|} &= \frac{\left(\begin{matrix} n - \sqrt{n}+ |\mcf{X}|-1 \\ |\mcf{X}|-1  \end{matrix} \right)}{\left(\begin{matrix} n +|\mcf{X}|-1  \\ |\mcf{X}|-1  \end{matrix} \right)} \\
%&= \prod_{k=1}^{\sqrt{n}} \frac{n-\sqrt{n}+k}{n-\sqrt{n}+|\mcf{X}|-1 + k} \\
&\geq   1 - \frac{\sqrt{n}(|\mcf{X}|-1)}{n-\sqrt{n}+|\mcf{X}|-1}
\end{align}
where the initial equality can be easily seen via a ``stars and bars'' proof, where the first bar must be chosen after the $\sqrt{n}$-th star.

The third and final technical result is  
\begin{align}
\sum_{\rho \in \mcf{P}_n(\mcf{X})} \frac{\rho(a)}{|\mcf{P}_n(\mcf{X})|} = \frac{1}{|\mcf{X}|} \label{eq:dmplsbale:tr3},
\end{align}
for all $ a\in \mcf{X}.$
Indeed, this result follows from combining
\begin{align}
\sum_{x^n \in \mcf{X}^n}    \frac{1}{|\mcf{T}^n_{(x^n)}||\mcf{P}_n(\mcf{X})|} =1 
\end{align}
and
\begin{align}
\sum_{x^n \in \mcf{X}^n}    \frac{1}{|\mcf{T}^n_{(x^n)}||\mcf{P}_n(\mcf{X})|} &=   \sum_{a \in \mcf{X}} \sum_{\rho \in \mcf{P}_n(\mcf{X})}  \frac{\rho(a)}{|\mcf{P}_n(\mcf{X})|} \\
&= |\mcf{X}|  \sum_{\rho \in \mcf{P}_n(\mcf{X})}  \frac{\rho(a)}{|\mcf{P}_n(\mcf{X})|} 
\end{align}
where the first equality is due to~\eqref{eq:dmplsbale:tr1}, and the second equality due to the inner sum must being equal for all $a \in \mcf{X}$ by symmetry .

Now with these technical results in tow, and recognizing that $\varepsilon(k)$ is monotonically decreasing with $k$ shows that the LHS of Equation~\eqref{eq:dmplsbale:nts} is less than or equal to
\begin{align}
&\sum_{\substack{ x^n \in \mcf{X}^{n}  ,\\ y \in \mcf{Y} , \\  \mcf{W}\subseteq \mcf{X}} }  \frac{s_{x^n}(y) u(x(1),\mcf{W}) }{ |\mcf{T}^n_{(x^n)}| |\mcf{P}_{n}(\mcf{X})|  } \varepsilon\left((n+|\mcf{X}|+1)\hat q^{(x^n)}(x(1))\right) \notag \\
&= \sum_{\substack{ x^n \in \mcf{X}^{n}  , \\  \mcf{W}\subseteq \mcf{X}} }  \frac{ u(x(1),\mcf{W}) }{ |\mcf{T}^n_{(x^n)}| |\mcf{P}_{n}(\mcf{X})|  } \varepsilon\left((n+|\mcf{X}|+1)\hat q^{(x^n)}(x(1))\right).
\end{align}
Next splitting up the summation by the type sets and which report needs to be summarizer, the above bounding can be continued with
\begin{align}
&\leq \sum_{\substack{\rho \in \mcf{P}_n(\mcf{X}), \\a \in \mcf{X}}}\sum_{  \substack{ x^n \in \mcf{X}^{n} : \\ \pi^{(x^n)} = \rho \\ x(1) = a }  }  \frac{ \sum_{  \mcf{W}\subseteq \mcf{X} }u(a,\mcf{W}) }{ |\mcf{T}^n_{(x^n)}| |\mcf{P}_{n}(\mcf{X})|  }  \varepsilon\left(n \rho(a) + |\mcf{X}|2^{-j} \right),
\end{align}
where the fact that $|\mcf{X}(y)| = |\mcf{X}|2^{-j}$ for a length $j$-summarizer has been used. 
To which we may apply our first technical result, Equation~\eqref{eq:dmplsbale:tr1}, yielding
\begin{align}
&= \sum_{a \in \mcf{X}}  \sum_{\rho \in \mcf{P}_n(\mcf{X})}  \rho(a)  \frac{ \sum_{  \mcf{W}\subseteq \mcf{X} }u(a,\mcf{W}) }{ |\mcf{P}_{n}(\mcf{X})|  }  \varepsilon\left(n \rho(a) + |\mcf{X}|2^{-j}  \right)
\end{align}
Which itself is
\begin{align}
&\leq \sum_{a \in \mcf{X}}  \sum_{\rho \in \mcf{P}_n(\mcf{X}): n \rho(a) < \sqrt{n}}  \frac{1}{\sqrt{n}}  \frac{ \sum_{  \mcf{W}\subseteq \mcf{X} }u(a,\mcf{W}) }{ |\mcf{P}_{n}(\mcf{X})|  }  \varepsilon\left( |\mcf{X}|2^{-j}  \right) \notag \\
&+ \sum_{a \in \mcf{X}}  \sum_{\rho \in \mcf{P}_n(\mcf{X}) }  \rho(a)  \frac{ \sum_{  \mcf{W}\subseteq \mcf{X} }u(a,\mcf{W}) }{ |\mcf{P}_{n}(\mcf{X})|  }  \varepsilon\left( \sqrt{n} \right) 
\end{align}
by $\varepsilon(k)$ being a monotonically increasing function of $k$ and recognizing that $n\rho(a) < \sqrt{n}$ also means that $\rho(a) \leq \frac{1}{\sqrt{n}}$.
Finally applying our second and third technical results, Equations~\eqref{eq:dmplsbale:tr2} and~\eqref{eq:dmplsbale:tr3}, yields 
\begin{align}
&\leq \sum_{\substack{a \in \mcf{X},\\ \mcf{W} \subseteq \mcf{X}} }\frac{u(a,\mcf{W})}{|\mcf{X}|}   \frac{(|\mcf{X}|-1)|\mcf{X}|}{n-\sqrt{n}+|\mcf{X}|-1}   \varepsilon\left( |\mcf{X}|2^{-j}  \right) \notag \\
&\hspace{10pt} + \sum_{\substack{a \in \mcf{X},\\ \mcf{W} \subseteq \mcf{X}}}   \frac{ u(a,\mcf{W}) }{ |\mcf{X}|  }  \varepsilon\left( \sqrt{n} \right) \notag \\
&= u^* \left( \frac{(|\mcf{X}|-1)|\mcf{X}|}{n - \sqrt{n} + |\mcf{X}|-1} \varepsilon(|\mcf{X}|2^{-j}) + \varepsilon(\sqrt{n}) \right) 
\end{align}

\end{IEEEproof}}{}

\end{document}